\documentclass[amssymb,
smallextended]{svjour3}       
\begin{document}

\title{Revisiting the Darmois and Lichnerowicz junction conditions
}


\author{Kayll Lake
}


\institute{Kayll Lake \at
              Department of Physics, \\Queen's University, \\Kingston, Ontario, Canada \\ K7L
3N6 \\
              Tel.: +1 613 533 2720\\
              Fax: +1 613 533 6463\\
              \email{lakek@queensu.ca}           
}

\date{Received: date / Accepted: date}

\maketitle

\begin{abstract}
What have become known as the ``Darmois" and ``Lichnerowicz" junction conditions are often stated
to be equivalent, ``essentially" equivalent, in a ``sense" equivalent, and so on. One even sees not infrequent reference to the ``Darmois-Lichnerowicz" conditions. Whereas the equivalence of these conditions is
manifest in Gaussian-normal coordinates, a fact that has been known for close to a century, this equivalence does not extend to a loose definition of ``admissible" coordinates (coordinates in which the metric and its first order derivatives are continuous). We show this here by way of a simple, but physically relevant, example. In general, a loose definition of the ``Lichnerowicz" conditions gives additional restrictions, some of which simply amount to a convenient choice of gauge, and some of which amount to real physical restrictions, away from strict ``admissible" coordinates. The situation was totally confused by a very influential, and now frequently misquoted, paper by Bonnor and Vickers, that erroneously claimed a proof of the equivalence of the ``Darmois" and ``Lichnerowicz" conditions within this loose definition of ``admissible" coordinates. A correct proof, based on a strict definition of ``admissible" coordinates, was given years previous by Israel. It is that proof, generally unrecognized, that we must refer to. Attention here is given to a clarification of the subject, and to the history of the subject, which, it turns out, is rather fascinating in itself.
\end{abstract}

\section{Introduction}
\label{intro}
\subsection{The Conditions}
\label{introc}
At a non-null boundary surface $\Sigma$ in spacetime
(we do not consider surface layers) consider the following
junction conditions\footnote{We include an absolute minimum of technical
complexity so as not to detract from the main points. Further, it it important to note that references throught this communication that were once obscure are now (for the most part) immediately available through the internet, though not always in translation.}:

\bigskip

\textit{Darmois 1927 }(D \cite{Darmois}): The continuity of the first
and second fundamental forms (the intrinsic metric and extrinsic curvature) \cite{eisenhart} across $\Sigma$.\footnote{It is to be emphasized that two different spacetimes are involved and their respective coordinates are unrelated prior to matching.}

\bigskip

\textit{Lichnerowicz 1955 }(L \cite{Lichnerowicz}): The continuity of
the metric and all first order partial derivatives of the metric
across $\Sigma$ in coordinates that traverse $\Sigma$ (such
coordinates being loosely referred to as ``admissible").\footnote{This is not exactly what Lichnerowicz said, but it \textit{is} what many references think he said. We start with this loose definition of ``admissible coordinates" and then rectify the definition in the section \textit{Admissible Coordinates} below.}

\subsection{Gaussian Normal Coordinates}
\label{introg}
In Gaussian-normal coordinates \cite{MTW} the spacetime is given by
\begin{equation}
ds^2=\pm d n^2+g_{i j}(n,x^k)dx^{i}dx^{j} \label{gauss}
\end{equation}
where $\Sigma$ is defined by $n=n_{0}$ where $n_{0}$ is a
constant. $\Sigma$ is timelike with ``+" and spacelike with ``-". Trajectories of constant $x^{i}$ are geodesics affinely parameterized by $n$. We suppose that the coordinates traverse $\Sigma$.
As is well known, the second fundamental forms of $\Sigma$ are simply
\begin{equation}\label{gaussk}
  K_{ij}=\frac{1}{2}\frac{\partial g_{i j}}{\partial n},
\end{equation}
and since the other partial derivatives of $g_{ij}$ lie in $\Sigma$, it immediately follows that
\begin{equation}\label{gaussdl}
  D \Leftrightarrow L
\end{equation}
and, it is important to note, \textit{in these coordinates} \cite{trivial}. At this point it is, perhaps, important to point out that the junction conditions are background \textit{geometrical} smoothness criteria and as such they can give rise to physically unreasonable situations even when satisfied \cite{unphysical}.
\subsection{Review of the Original Literature}
\label{intror}
In \cite{Darmois} page 29 Darmois says

\bigskip

``\textit{The conditions that we will meet, and whose importance is very high if we want to understand the interdependence of the masses and the field, were introduced by Schwarzschild. We will put as a condition that there is a system of coordinates in which the $g_{\alpha \beta}$ and their first derivatives are continuous functions even at the crossing of the border}"

\bigskip

What is interesting is that not only did Darmois state a weak form of the L conditions, he attributes them to Schwarzschild. Whereas Darmois quotes both of Schwarszchild's papers (the exterior and interior solutions) it is really the interior solution paper \cite{schw} that makes the point clear. In any event, whereas it might be argued that Schwarzschild was considering a special case, Darmois goes on to say

\bigskip

``\textit{These conditions are not invariant under an arbitrary change of variables and in reality can present difficulties. It is not always obvious whether the variables employed for solving the two problems, interior and exterior, allow the implementation of these conditions}"

\bigskip

Darmois then points out that it is always possible to introduce Gaussian normal coordinates (\ref{gauss}), states that the first fundamental form is $g_{ij}$ and that the second fundamental form is (\ref{gaussk}). He then says

\bigskip

``\textit{the} [continuity of the] \textit{first and second fundamental forms of the border represent the common boundary}"

\bigskip
and

\bigskip
``\textit{This condition is invariant, and can now be expressed, without otherwise effecting the change of coordinates, in any system.}"

\bigskip

And so, whereas Darmois used Gaussian normal coordinates to introduce the idea of the continuity of the first and second fundamental forms, he clearly realized that these conditions do not depend on the coordinates of the enveloping spacetimes.

\bigskip

In \cite{Lichnerowicz} Lichnerowicz devotes a chapter (III) to junction conditions. (The idea of ``admissible" coordinates can be found early in chapter I.) However, essentially the same material can be found in his earlier work \cite{Lichnerowicz1}. There, on page 31, in a section called ``\textit{The junction conditions of Schwarzschild}" one can find essentially everything except precise details  concerning ``admissible" coordinates.\footnote{This is explained by Lichnerowicz in \cite{Lichnerowicz88}.} \footnote{The idea that ``admissible" coordinates could be constructed by way of the introduction of coordinate changes fixed by the matching procedure was, of course, not new even at that time. This is the procedure used by Oppenheimer and Snyder \cite{os}, and on a mathematically equivalent problem by Einstein and Straus \cite{es}. } Darmois was Lichnerowicz's supervisor, and one can find a reference to \cite{Darmois} in the Introduction of \cite{Lichnerowicz1} and in \cite{Lichnerowicz} one can find a long introduction by Darmois. It remains a fact of history that Lichnerowicz, fully aware of Darmois' suggestion to use the first and second fundamental forms, did not pursue this approach. The reason, most certainly, is the fact that Lichnerowicz was interested primarily in \textit{global} problems, even in these early days, and ``admissible" coordinates (when properly defined) suited his needs. Of course, global problems have evolved into an important area of study \cite{global}.

\subsection{The Early Impact}
\label{introi}
The work of Lichnerowicz \cite{Lichnerowicz} was soon well known. In the influential book by Synge \cite{Synge}, one can find reference to \cite{Lichnerowicz} on page 1. Further, on page 40, Synge says

\bigskip

``\textit{Much of the work done on junction conditions prior to the introduction of admissible coordinates by Lichnerowicz is mathematically obscure.}"

\bigskip

The work of Darmois \cite{Darmois} was not well known, at least outside of France,\footnote{Within France, the equivalence of the Darmois and (correctly formulated) Lichnerowicz conditions were well known (Jos\'{e} Senovilla, private communication).} until the middle 1960's. As late as 1962, Beckedorff \cite{Beckedorff} solved the Oppenheimer - Snyder problem using the continuity of the first and second fundamental forms without reference to Darmois. The D conditions, with reference to Beckedorff, but not to Darmois, are stated explicitly by Misner and Sharp in 1964 \cite{MS}\footnote{For a related work that cites \cite{Lichnerowicz} and \cite{MS} but not Darmois see \cite{Bel}.}. In the fall of 1965 Israel submitted his now famous work \cite{Israel} in which he developed the D conditions and extended the analysis to the study of surface layers based on discontinuities in the second fundament forms. Israel gives an extensive review of previous work but with no mention of Darmois.\footnote{Israel was simply not aware of the paper by Darmois (Werner Israel, private communication).} The first reference to Darmois, outside of France, would appear to be the paper by Cocke \cite{Cocke}. The fact that Cocke was at the Institut Henri Poincar\'{e} might help explain this. After the paper by Cocke, there was a growing list of references to Darmois. Because of the importance of surface layers, one now finds frequent reference to the ``Darmois-Israel" conditions.\footnote{It must be noted that Darmois never considered discontinuities in the second fundamental forms.}

\section{The Paper by Bonnor and Vickers}
\label{introbv}
In 1981 a very influential paper by Bonnor and Vickers \cite{bv} appeared \cite{Web}. Whereas this paper was primarily interested in the junction conditions of O'Brien and Synge, which we do not discuss here, it is known primarily for the claimed proof of the
\textit{equivalence} of D and L. Their argument can be summarized as
follows:

\bigskip

Given D one can introduce Gaussian-normal coordinates and
so L is satisfied. Let us write this here as D $\Rightarrow$ L. To
establish L $\Rightarrow$ D they argued that if L is satisfied then so is D.

\bigskip

This brief ``proof" is not adequate to establish the equivalence of D and L. As regards L $\Rightarrow$ D, one must address the issue as to whether L is simply assumed in some set of coordinates, and whether or not the set of conditions L are in fact the same as the conditions D. For example, the conditions D could be a subset of the conditions L in which case the conditions can not be said to be equivalent. As regards D $\Rightarrow$ L, aside from the fact that the introduction of Gaussian-normal coordinates that traverse $\Sigma$ may be impractical, one needs precise conditions under which, given D, coordinates that traverse $\Sigma$ can be constructed in which the metric is differentiable at $\Sigma$. This is discussed below.

\subsection{The Impact}
\label{introbvi}
 An inexhaustive search of the literature reveals the following: (i)
An outright claim that D and L are equivalent (citing \cite{bv}
for a proof) is given in, for example, \cite{SS}, \cite{GD},  \cite{wiltshire},
\cite{Oliwa}, \cite{gonna}, \cite{maccallum}, \cite{beruni},
\cite{fayos}, \cite{wiltshire1}, \cite{clarke}, and
\cite{herrera}. (ii) A claim of equivalence (not citing \cite{bv})
is in \cite{hellaby}. (iii) A claim that D and L are
``essentially" equivalent is in \cite{pereira}. (iv) Sometimes D
and L have been discussed without the alleged equivalence given in \cite{bv}, for
example in \cite{fayos1}, \cite{mansouri} and \cite{mars}. (v)
More cautionary statements that D are more ``reliable" than L can
be found in \cite{hellaby1}. (vi) Also cautionary statements that
D, ``according to \cite{bv}", are equivalent to L can be found in \cite{Bhar}, \cite{Bhar1} and \cite{hellaby3}. (vii) There are statements in which I think that
the author is saying D is equivalent to L  according to \cite{bv},
but I am not sure, as for example in \cite{maccallum1}. (viii)
There are references that paraphrase \cite{bv}, for
example \cite{book2}  page 47, but do not state that the L conditions,
away from Gaussian-normal coordinates, are not necessarily
equivalent to the D conditions. (ix) There are references that attribute the L conditions to \cite{bv}, for example \cite{PAP}, \cite{PPM} and \cite{MP}. (x) There are references that refer to the Darmois-Lichnerowicz conditions \cite{JK}. (xi) There is even a reference that claims that the book by Lichnerowicz \cite{Lichnerowicz} was coauthored by Darmois \cite{CGR}!

\section{An Example}
\label{example}
Here we consider the classic problem of the junction of a spherically symmetric static (not necessarily perfect) fluid
onto the Schwarzschild vacuum. To compare the D and L conditions, we work out the junction problem in Gaussian normal coordinates and then in curvature coordinates.
\subsection{Gaussian Normal Coordinates}
\label{exampleg}
We start with the line element
\begin{equation}\label{gaussl}
  ds^2 = dr^2+A(r)d\Omega^2_2+B(r)dt^2
\end{equation}
where $d\Omega_2^2$ represents the line element of a two-sphere ($d\theta^2+\sin(\theta)^2d\phi^2$). We take the coordinates to be comoving and take an energy-momentum tensor to be of the form
\begin{equation}\label{energy}
T^{\alpha}_{\beta}=diag[p(r),P(r),P(r),-\rho(r)].
\end{equation}
By writing out the Einstein tensor $G^{\alpha}_{\beta}$ we see that $P$ and $\rho$ involve second order derivatives of the functions $A$ and $B$ but $p$ does not, it involves only first order derivatives. We conclude that $p$ is necessarily $0$ at $\Sigma$ which can be taken to be $r=constant = r_{\Sigma} >0$. This allows us to write $B$ in terms of $A$ for $r \geq r_{\Sigma}$.
A fundamental property of the metric (\ref{gaussl}) is the effective gravitational mass, the invariant properties of which were first explored by Hernandez and Misner
\cite{hm} who wrote the function in the form
\begin{equation}
m(r)= \frac{A^{3/2}}{2}\mathcal{R}_{\theta \phi}^{\; \; \;\; \theta
\phi} , \label{mass}
\end{equation}
where $\mathcal{R}$ is the Riemann tensor. See also \cite{mass}.  From (\ref{gaussl}) and (\ref{mass}) we find
\begin{equation}\label{explicit}
  m(r)=\frac{4A-A'^2}{8 \sqrt{A}}
\end{equation}
and so $m(r)$ is also continuous across $\Sigma$ with $m=m(r_{\Sigma})$ for $r \geq r_{\Sigma}$ \cite{explicit}. This completes the junction problem in Gaussian normal coordinates.

\subsection{Curvature Coordinates}
\label{examplec}
Let us start with the D conditions. The exterior Schwarzschild
vacuum, in terms of exterior curvature coordinates
$(\textsf{r},\theta,\phi,T)$, is of course given by
\begin{equation}
ds^2 = \frac{d\textsf{r}^2}{1-\frac{2 m}{\textsf{r}}} +
\textsf{r}^2d\Omega^2_2-(1-\frac{2 m}{\textsf{r}})dT^2.
\label{vacuum}
\end{equation}
The interior line element, in
terms of interior comoving curvature coordinates
$(r,\theta,\phi,t)$, can be taken to be
\begin{equation}
ds^2=\frac{dr^2}{1-\frac{2m(r)}{r}}+r^2d\Omega^2_2-e^{2\Phi(r)}dt^2.
\label{standard}
\end{equation}
The coordinates in (\ref{gaussl}) and in (\ref{standard}) are entirely distinct.
Without loss in generality we can take
\begin{equation}
ds^{2}_{\Sigma}=R(\tau)^2d\Omega^2_2-d\tau^2 \label{surface}
\end{equation}
where $\tau$ is the proper time on $\Sigma$.

\bigskip

The continuity of the intrinsic metric
associated with $\Sigma$ ensures that the continuity of $\theta$
and $\phi$ in metrics (\ref{vacuum}) and (\ref{standard}) is
allowed and that the history of the timelike boundary surface $\Sigma$ is
given by
\begin{equation}
R(\tau)=r_{\Sigma} (>2m(r_{\Sigma}))= \textsf{r}_{\Sigma} (> 2 m).
\label{history}
\end{equation}
The continuity of the extrinsic curvature component $K_{\tau \tau}$ gives us
\begin{equation}\label{ktt}
  \Phi^{'}\bigm|_{\Sigma} = \frac{m}{\textsf{r}(\textsf{r}-2m)}\bigm|_{\Sigma}
\end{equation}
 so that along with the source equation for $\Phi$ (the generalized Tolman
- Oppenheimer - Volkoff equation) we have
\begin{equation}
p(r_{\Sigma})=0 \label{psurf}
\end{equation}
where $p$ is, again, the isotropic pressure. We can take (\ref{psurf})
(with (\ref{history})) as the definition of the boundary $\Sigma$.
The continuity of the extrinsic curvature components $K_{\theta
\theta}$ and $K_{\phi \phi}$, along with (\ref{history}), gives
\begin{equation}
m(r_{\Sigma})= m. \label{msurf}
\end{equation}
Finally, the correct rigging of the 4-normals to $\Sigma$ (interior to
exterior) is verified by the continuity of the trace of the
extrinsic curvature across $\Sigma$. Equations (\ref{history}),
(\ref{psurf}) and (\ref{msurf}) then constitute the solution to
the D junction problem for the case of a static spherically
symmetric fluid joined onto vacuum. We see that the D conditions are invariant to the change in coordinates,
as they must be. Even in this simple case, it is to be noted that the D conditions are not entirely trivial to execute.

\bigskip

To explore the L conditions let us start with the
\textit{assumption} that the coordinates used in (\ref{vacuum})
and (\ref{standard}) are admissible in the weak sense (\textit{i.e.} there is no
distinction between $(t,T)$ and $(\textsf{r},r)$ and the 4-metric
components are $C^1$ at $\Sigma$). Then, in addition to the
previous conditions, when applied to $(t,T)$ the L conditions also
give
\begin{equation}
e^{2\Phi(r_{\Sigma})}=1-2\frac{m}{\textsf{r}_{\Sigma}},
\label{tgauge}
\end{equation}
(the derivative requirement already appearing in (\ref{ktt})) and when applied to $(\textsf{r},r)$ they give the further
condition
\begin{equation}
m^{'}(r_{\Sigma})=\rho(r_{\Sigma})=0, \label{rhojump}
\end{equation}
where $\rho$ is, again, the comoving energy density.   Condition (\ref{tgauge}) is a forced, but convenient, choice of gauge that we discuss below. The condition
(\ref{rhojump}) is a physical restriction. There \textit{are} solutions that satisfy (\ref{rhojump}) \cite{rho} and so for these, with (\ref{tgauge}), the L conditions are satisfied. We conclude that D does not imply L for these solutions.
\subsection{An aside on Gauge Conditions}
\label{examplegc}
It is well-known that the evolution of null geodesics in the spacetime (\ref{standard}) is governed by
\begin{equation}\label{effective}
  \frac{\dot{r}^2}{l^2}=\frac{1}{r^2} \left(1-\frac{2m(r)}{r} \right) \left(\frac{\mathcal{B}(r)^2}{b^2}-1\right)
\end{equation}
where $^{.} = d/d \lambda$, $\lambda$ being an affine parameter, $b^2 = \gamma^2/l^2$ where $\gamma$ (the ``energy") is associated with the Killing vector $\delta^{\alpha}_{t}$ and $l$ (the ``angular momentum") is associated with the Killing vector $\delta^{\alpha}_{\phi}$, and the ``potential impact parameter" $\mathcal{B}(r)^2$ is given by
\begin{equation}\label{potential}
  \mathcal{B}(r)^2 = \frac{r^2}{e^{2 \Phi(r)}}.
\end{equation}
With the aide of condition (\ref{tgauge}) we see that $\mathcal{B}^2$ is continuous through $\Sigma$ and with the addition of (\ref{ktt}) we have $(\mathcal{B}^2)^{'}$ continuous through $\Sigma$. The gauge condition (\ref{tgauge}) therefore provides a convenient choice when studying, for example, transparent spherical gravitational lenses.

\bigskip

For the complete Schwarzschild solution, the interior is given by
\begin{equation}\label{schwi}
  ds^2=\frac{dr^2}{1-\frac{r^2}{R^2}}+r^2 d\Omega^2_2-(a-b\sqrt{1-\frac{r^2}{R^2}})^{2}dt^2
\end{equation}
where $a$ and $b$ are constants as is $\rho$ where $8 \pi \rho/3 = 1/R^2$. The lensing properties of the complete Schwarzschild solution have been studied \cite{schwl}. If $r_{\Sigma}$ exists, it is given by
\begin{equation}\label{intsigma}
  3\sqrt{1-\frac{r^{2}_{\Sigma}}{R^2}}=\frac{a}{b}.
\end{equation}
In the first edition of the \textit{Exact Solutions} book \cite{book1}, the gauge condition (\ref{tgauge}), which amounts to $b=1/2$ in this case, is applied. This is also true of the book by Griffiths and Podolsk\'{y} \cite{GP}. In the second edition of the \textit{Exact Solutions} book \cite{book2}, there is no mention of $r_{\Sigma}$ and no gauge condition is applied. The point is that the gauge condition \ref{tgauge} is convenient, but not required.

\section{Admissible Coordinates}
\label{admissible}
In the foregoing, an essential element to the Lichnerowicz approach has been lost. In \cite{Lichnerowicz}, on page 5, Lichnerowicz says

\bigskip

``\textit{We demand that in the intersection between the domains of two admissible coordinate systems, the second derivatives of the coordinate transformation must be piecewise $C^2$ functions.}"

\bigskip

At first sight, this might appear a little strong. The fact that Lichnerowicz frequently uses the requirement that \textit{derivatives} of the coordinate transformation must be piecewise $C^2$ functions perhaps confuses the issue.\footnote{This confusion persists. To quote from a recent paper \cite{lapiedra} ``\textit{the term ``admissible" designates a coordinate system of a $C^2$ class (atlas) manifold structure describing the space-time}". } The original statement is correct. If D \textit{has been established}, then one can prove (see \cite{israel2}, \cite{Israel} and more recently \cite{mars}) that there always exists coordinates, by way of $C^1$ (piecewise $C^3$) transformations, that traverse $\Sigma$ such that the metric is differentiable at $\Sigma$. Call these coordinates $\mathcal{C}$. Israel's transparent notation for $\mathcal{C}$ is\footnote{Curiously, Synge \cite{Synge}, on page 2, refers to the $C^3$ condition as \textit{``not the important thing"}, but it is.}
\begin{equation}\label{israel}
  g_{\alpha \beta} = (C^1_{\Sigma},C^3).
\end{equation}
Further (see the same references above), all coordinates obtained from $\mathcal{C}$ via $C^2$ (piecewise $C^4$) transformations, in Israel's notation
\begin{equation}\label{c4}
  (C^2_{\Sigma},C^4),
\end{equation}
form a set of coordinates, say $\mathcal{S}$, that we call ``admissible" or ``natural" coordinates exactly as stated by Lichnerowicz in \cite{Lichnerowicz88}.\footnote{An explanation of these conditions, in English, was given by Lichnerowicz at the 1957 Chapel Hill Conference. Unfortunately, his lecture remained rather difficult to get until relatively recently. See \cite{Lichnerowicz57}.  } If the conditions L are carried out in $\mathcal{S}$ then clearly D and L are equivalent. The flaw in the proof by Bonnor and Vickers \cite{bv} is the lack of a requirement of \textit{smoothness} which is \textit{the} essential feature of L. The flaw in the example given above is that curvature coordinates are simply not ``admissible", they were simply \textit{assumed} to be.

\section{Conclusions}
\label{conclusions}
As should be well known, the D and L conditions are equivalent in Gaussian normal coordinates and in coordinates related to Gaussian normal coordinates by $(C^2_{\Sigma},C^4)$ transformations (such coordinates being referred to as ``natural" or ``admissible" coordinates $\mathcal{S}$). Whereas the D conditions can be examined in any set of enveloping coordinates, albeit with some effort\footnote{In an attempt to alleviate the complications of following the Darmois-Israel approach, the package \textit{GRJunction} \cite{musgrave1}, \cite{musgrave2} was developed by Peter Musgrave to run with \textit{GRTensorII}. Unfortunately, due to my own negligence, \textit{GRJunction} was never in wide distribution, despite the very wide spread use of \textit{GRTensorII}. This has now changed. \textit{GRTensorII} has been updated to \textit{GRTensorIII}, thanks to Peter's efforts, and an update to \textit{GRJunction} has been completed and is now available \cite{GRTensorIII 2.0}.}, the L conditions are restricted to ``natural" or ``admissible" coordinates in the strong sense of $\mathcal{S}$. It is fitting to end with a quotation from \cite{Israel}. When reflecting on section 7 of his paper, Israel says

\bigskip

``\textit{The formulas of this Section, despite their simple appearance, are actually of limited utility, since natural} [admissible] \textit{coordinates are seldom the most convenient for handling a problem in practice.}"

\bigskip

\textit{Acknowledgments.} This work was supported by a grant from the Natural Sciences and Engineering Research Council of Canada. It is a pleasure to thank Bill Ballik, Werner Israel, Dmitri Lebedev, Peter Musgrave, Eric Poisson and Jos\'{e} Senovilla for discussions.


\begin{thebibliography}{}
\bibitem{unphysical} Following on \cite{trivial}, and considering the case of Robertson-Walker geometries, we see that for any (spacelike) surface of comoving proper-time, we only need $a$ and its derivative to be continuous. This allows instantaneous ``phase changes" which are unjustifiable without further physics describing the situation.
\bibitem{Web}At the time of writing, this paper has 96 citations via the Web of Science.
\bibitem{explicit} Setting $m(r)=m$ in (\ref{explicit}) it does not follow that $A$ can in fact be expressed in terms of elementray functions. This in no way affects the argument given here.
\bibitem{rho} Consider, for example, $m(r)=-r^4+a r^3$ with $0 < a < (32/9)^{(1/3)}$. We have $4 \pi \rho(0)=3a$, $\rho|_{\Sigma}=0$ for $r_{\Sigma}=3a/4$ and $\Phi$ is a monotone increasing function of $r$ with a regular minimum at $r=0$ and subject to the boundary condition $\Phi^{'}_{\Sigma}=6a^2/(32-9a^3)$.
\end{thebibliography}
\end{document}